\documentclass[conference]{IEEEtran}
\IEEEoverridecommandlockouts
\usepackage{cite}
\usepackage{amsmath,amssymb,amsfonts}
\usepackage{multirow}
\usepackage{algorithmic}
\usepackage{graphicx}
\usepackage{textcomp}
\usepackage{xcolor}
\usepackage{booktabs}
\usepackage{marvosym}
\def\BibTeX{{\rm B\kern-.05em{\sc i\kern-.025em b}\kern-.08em
    T\kern-.1667em\lower.7ex\hbox{E}\kern-.125emX}}
\begin{document}

% \title{DiffSR: Two-Stage Diffusion Model for Reconstructing Radar Reflectivity via Satellite Data\\
% }

\title{DiffSR: Learning Radar Reflectivity Synthesis via Diffusion Model from Satellite Observations \\
}

\author{\IEEEauthorblockN{1\textsuperscript{st} Xuming He}
\IEEEauthorblockA{\textit{Open Science Lab} \\
\textit{Shanghai AI Laboratory}\\
Shanghai, China \\
hexuming773@gmail.com}
\and
\IEEEauthorblockN{2\textsuperscript{nd} Zhiwang Zhou}
\IEEEauthorblockA{\textit{Open Science Lab} \\
\textit{Shanghai AI Laboratory}\\
Shanghai, China \\
zhouzhiwang@pjlab.org.cn}
\and
\IEEEauthorblockN{3\textsuperscript{rd} Wenlong Zhang\textsuperscript{\Letter}}
\IEEEauthorblockA{\textit{Open Science Lab} \\
\textit{Shanghai AI Laboratory}\\
Shanghai, China \\
zhangwenlong@pjlab.org.cn}
\and
\IEEEauthorblockN{4\textsuperscript{th} Xiangyu Zhao}
\IEEEauthorblockA{\textit{Department of Computing} \\
\textit{Hong Kong Polytechnic University}\\
HongKong SAR, China \\
22123675r@connect.polyu.hk}
\and
\IEEEauthorblockN{5\textsuperscript{th} Hao Chen}
\IEEEauthorblockA{\textit{Open Science Lab} \\
\textit{Shanghai AI Laboratory}\\
Shanghai, China \\
chenhao1@pjlab.org.cn}
\and
\IEEEauthorblockN{6\textsuperscript{th} Shiqi Chen}
\IEEEauthorblockA{\textit{Shanghai Marine Meteorological Center} \\
\textit{Shanghai Meteorological Service}\\
Shanghai, China \\
onlycsq@163.com}
\and
\IEEEauthorblockN{7\textsuperscript{th} Lei Bai\textsuperscript{\Letter}}
\IEEEauthorblockA{\textit{Open Science Lab} \\
\textit{Shanghai AI Laboratory}\\
Shanghai, China \\
baisanshi@gmail.com}
\thanks{\textsuperscript{\Letter}Corresponding author}
}

\maketitle

\begin{abstract}
Weather radar data synthesis can fill in data for areas where ground observations are missing. Existing methods often employ reconstruction-based approaches with MSE loss to reconstruct radar data from satellite observation. However, such methods lead to over-smoothing, which hinders the generation of high-frequency details or high-value observation areas associated with convective weather. To address this issue, we propose a two-stage diffusion-based method called DiffSR. We first pre-train a reconstruction model on global-scale data to obtain radar estimation and then synthesize radar reflectivity by combining radar estimation results with satellite data as conditions for the diffusion model. Extensive experiments show that our method achieves state-of-the-art (SOTA) results, demonstrating the ability to generate high-frequency details and high-value areas. 
% Radar composite reflectivity is widely used to identify severe convective weather and is closely linked to precipitation-related phenomena.
% However, Weather radar data coverage is often limited by natural factors like terrain. Geostationary satellites, in contrast, provide full-coverage, real-time observational data to fill these gaps. 
% This study proposes a two-stage conditional diffusion model to reconstruct radar composite reflectivity from satellite data. 
% Our model shows notable improvements in meteorological metrics such as CSI, FAR, and POD, and image reconstruction metrics like LPIPS and SSIM, indicating a superior ability to capture detailed features. 
% Furthermore, the independent training of the model’s two stages allows it to accept inputs from various first-stage models, enhancing its universality.
\end{abstract}

\begin{IEEEkeywords}
Radar composite reflectivity, conditional diffusion model, Radar reconstruction
\end{IEEEkeywords}

\section{Introduction}
Weather radar data, particularly radar composite reflectivity (REFC), is essential for identifying severe convective weather events, such as thunderstorms, tornadoes, and heavy precipitation \cite{wang2021characteristic, borga2022rainfall, binetti2022use}. This information allows meteorologists to track weather patterns in real-time and issue timely warnings to reduce the high impact on people's lives and economic development \cite{xu2021meteorological,rigo2021evolution,nastos2021risk}. However, conventional radar systems rely heavily on ground-based stations, leading to significant gaps in coverage in remote regions, mountainous areas, and oceans~\cite{meyers2012mountain}. 
To overcome these limitations, data-driven models are being developed to utilize satellite data for radar data reconstruction~\cite{yu2023radar, yang2022radar,wang2023maformer, stock2024srvit,duan2021reconstruction}, offering a promising approach for improving the accuracy of severe weather detection.

Recently, deep learning technology has achieved rapid development in image super-resolution~\cite{zhang2019ranksrgan,zhang2021ranksrgan,zhang2022closer,zhang2024real}, restoration~\cite{chen2023hat,cao2024grids,genlv}, and the enhancement of spatial resolution of meteorological satellites~\cite{hu2018deconvolution,hu2019spatial,li2019spatial}.
Among various deep learning models, UNet-based \cite{yu2023radar, yang2022radar} and Transformer-based \cite{wang2023maformer, stock2024srvit} architectures have been extensively applied to the task of radar reflectivity reconstruction from satellite observations. UNet-based models are particularly effective in capturing spatial features due to their encoder-decoder structure, while Transformer-based models leverage attention mechanisms to model complex spatial and temporal dependencies in weather data.
However, the existing approaches that employ MSE loss for optimization inevitably tend to generate over-smoothing results. Consequently, \textbf{\textit{this issue leads to two limitations, including the failure to capture high-frequency details at a fine granularity and the challenges of producing high-value observation areas that represent strong convective weather.}}

% However, both types of models suffer from limitations, particularly in generating radar images that lack sharpness and fine-grained details. 
% These issues highlight the need for more advanced approaches that can enhance the clarity of reconstructed radar data.

In this work, we propose a novel two-stage diffusion-based framework (i.e., DiffSR) to overcome the limitations. The first stage utilizes Vision Transformer (ViT)~\cite{dosovitskiy2020image} to produce an initial coarse estimation of radar reflectivity from satellite observations, In the second stage, a diffusion model is introduced to refine the initial estimation by using the combination of initial estimation and satellite data as conditions. 
% This conditional diffusion process iteratively recovers fine-grained details. 
By leveraging the strengths of ViTs for generating initial global estimation and diffusion models for detail refinement, DiffSR offers a robust solution for producing high-quality radar reflectivity reconstructions from satellite observations.

\section{Method}
\begin{figure*}[ht]
  \centerline{\includegraphics[width=0.9\textwidth]{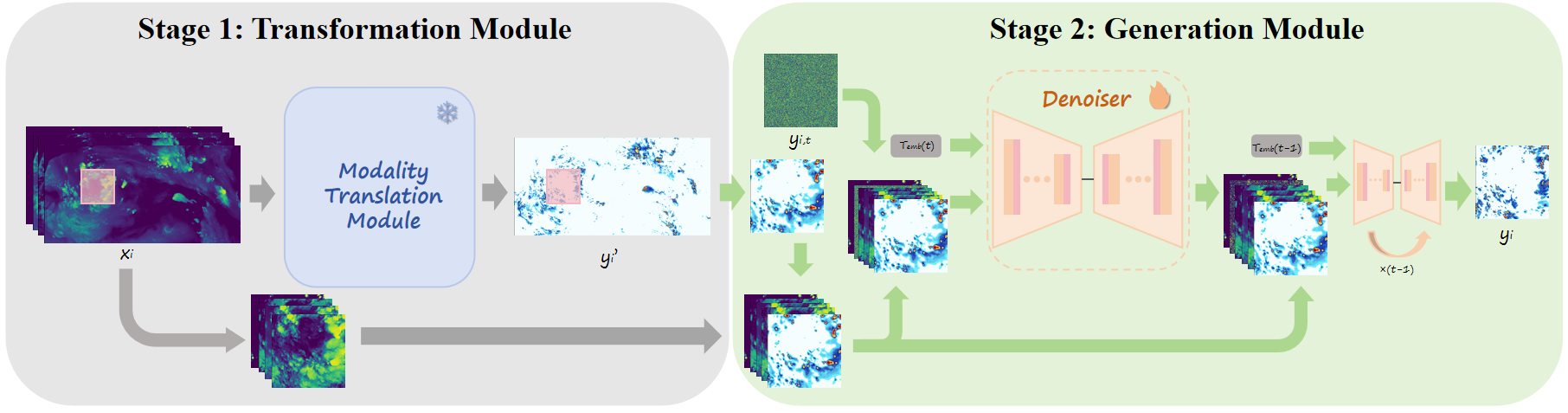}} 
  \caption{The two-stage pipeline of DiffSR. First, we use a fixed modality transformation module (MTM) to generate radar reconstruction image $y_i'$ from satellite and lightning conditions $x_i$ on a image-level. Then, we employ a conditional diffusion model to iteratively denoise pure Gaussian noise with the patch-level combination of $y_i'$ and $x_i$ as conditions and finally get the reconstruction results $y_i$.}
  \label{architecture}
\end{figure*}
% \subsection{Preliminaries}
% \subsubsection{Problem Formulation}
\subsection{Motivation and Framework}
Weather radar data synthesis aims to generate a radar data estimation from satellite observation, which can be defined as
\begin{equation}
    \mathcal{T}: \mathbb{X} \to \mathbb{Y}, \,\mathbb{X}=\{x_i\}_{i=1}^N \,\mathbb{Y}=\{y_i\}_{i=1}^N
\end{equation}
where $x_i$ represents the satellite observation and $y_i$ denotes the weather radar observation.

In this work, we exploit a powerful generative diffusion model to solve the problem of radar data synthesis. 
The generative diffusion model~\cite{saharia2022image,wang2024sam} has demonstrated its effectiveness in conditional image generation by using condition inputs, such as edge, segmentation map, and specific image. This provides a potential solution for high-quality radar data synthesis. 
However, directly using satellite data as a condition may lead to inaccurate results due to the powerful generative capabilities of diffusion models. 
To address this issue, we introduce a transformation module to first transform the satellite data into a radar estimation $y'$ on an image level. 
% that includes the observation contours and high-value areas.
Subsequently, we use both the patch-level satellite data and the radar estimation as conditions to generate radar composite reflectivity for the diffusion model.

\subsection{Transformation Module}
% In the first stage, we aim to generate radar composite reflectivity $y_i'$ with well-defined contours but less detailed information based on satellite data. 
In the first stage, we aim to generate radar composite reflectivity $y_i'$ with well-defined contours from satellite observation data $x_i$ on a image-level.
As shown in Figure~\ref{architecture}, we utilize image-level data directly to train the transformation module. This strategy can effectively capture global relationships within the full satellite image observation.
% This serves as a reliable condition for the generation module, allowing it to focus on refining details without being affected by noise corruption. The model for the first stage can be represented as:
\begin{equation}
\mathcal{TM}(x_i)=y_i'
\end{equation}
Specifically, we train a transformation module (TM) with weighted loss from \cite{hilburn2020development} to balance high reflectivity values, which are rare, against the more common but smaller values.
\begin{equation}
\mathcal{L}_e=\frac{1}{m}\sum_{i=1}^m\exp(w_0t_i^{w_1})\cdot(y_i'-t_i)^2,
\end{equation}
where $t$ and $y'$ are the ground truth and reconstructed values, $m$ is the number of pixels. The pre-trained transformation module performs as a condition preprocessing for the generation module.
% and it will be discarded during inference due to its inability to produce fine details.

% \subsubsection{Diffusion Models}

\subsection{Generation Module}
\textbf{Preliminary: Diffusion Model.} Inspired by \cite{saharia2022image,shang2024resdiff}, we use a conditional denoising diffusion model in the second stage to reconstruct composite radar reflectivity. 

The conditional diffusion model generates a target $y_i$ over $T$ consecutive steps starting from pure Gaussian noise. At each time step, the model takes in a time step vector and a condition $y_i'$, iteratively denoises the image output from the previous step according to a learned conditional distribution $p(y_{t-1}|y_t,y_i')$, and ultimate obtaining $y_0 \sim p(y|y')$.

The forward diffusion process gradually adds random noise to the target $y_0$, with the noise intensity controlled by a variance schedule $\beta_1$, $\beta_2$, ..., $\beta_t$ for each time step. This is achieved through a fixed Markov chain of length $T$, generating a sequence of data $y_1, y_2, ..., y_T$.
\begin{equation}
q(y_t|y_{t-1})=\mathcal{N}(y_t|\sqrt{1-\beta_t}y_{t-1}, \beta_tI)\label{eq1}
\end{equation}

If the $\beta_t$ values are set small enough, and the total time step $T$ is sufficiently large, the properties of the Markov chain enable us to connect $q(y)$ with $\mathcal{N}(0, 1)$. We can then obtain the distribution of $y_t$ using the following equation:
\begin{equation}
q(y_t|y_0)=\mathcal{N}(y_t|\sqrt{\bar{\alpha_t}}y_0, (1-\bar{\alpha_t})I)\label{eq2},
\end{equation}
where $\alpha_t=1-\beta_t$, and $\bar{\alpha_t}=\prod_{i=1}^t a_i$. 

In the reverse process, we aim to estimate $q(y_{t-1}|y_t)$ to remove the noise added in the forward process. However, $q(y_{t-1}|y_t)$ can't be estimated directly. Instead, we train a neural network $p_\theta$, conditioned on $y_t$ and conditional data $y'$, to approximate the distribution of $y_{t-1}$.
\begin{equation}
p_\theta(y_{t-1}|y_t, y')=\mathcal{N}(y_{t-1}|\mu_\theta(y_t, y', \bar{\alpha_t}), \sigma_t^2I)\label{eq3}
\end{equation}

Following \cite{ho2020denoising}, the mean of $p_\theta(y_{t-1}|y_t, y')$ can be parameterized as follows:
\begin{equation}
\mu_\theta(y_t, y', \bar{\alpha_t})=\frac{1}{\sqrt{\alpha_t}}\Big( y_t- \frac{1-\alpha_t}{\sqrt{1-\bar{\alpha_t}}}\epsilon_\theta(y_t, y', \bar{\alpha_t})\Big)\label{eq4}
\end{equation}

Following this parameterization, we can get the distribution of $y_{t-1}$ and finish the reverse process.
\begin{equation}
y_{t-1}=\frac{1}{\sqrt{\alpha_t}}\Big( y_t- \frac{1-\alpha_t}{\sqrt{1-\bar{\alpha_t}}}\epsilon_\theta(y_t, y', \bar{\alpha_t})\Big)+\sqrt{\beta_t},
\label{eq5} 
\end{equation}
where $\epsilon\sim\mathcal{N}(0,1)$.
\begin{table*}[ht]
\centering
\caption{Model Performance Comparison for Radar Reflectivity Reconstruction (The Model denoted with $^*$ is trained by ourselves)}
\label{comparison}
\begin{tabular}{c|c|c|c|ccc|ccc|ccc}
\hline
\multirow{2}{*}{\centering\textbf{Model}} & \multirow{2}{*}{\textbf{$\downarrow$ RMSE}} & \multirow{2}{*}{\textbf{$\downarrow$ LPIPS}} & \multirow{2}{*}{\textbf{$\uparrow$ SSIM}} & \multicolumn{3}{c|}{\textbf{$\uparrow$ CSI-15}} &
\multicolumn{3}{c|}{\textbf{$\uparrow$ CSI-35}} &
\multicolumn{3}{c}{\textbf{$\uparrow$ CSI-50}}\\
\cline{5-13}
& & & & POOL1 & POOL4 & POOL8 & POOL1 & POOL4 & POOL8 & POOL1 & POOL4 & POOL8 \\
\hline
UNet & 3.61 & 0.359 & 0.780 & 0.448 & 0.428 & 0.441 & 0.221 & 0.275 & 0.321 & 0.086 & 0.111 & 0.130\\
SRViT & \textbf{3.40} & 0.320 & 0.802 & 0.473 & 0.478 & 0.490 & 0.243 & 0.318 & 0.362 & 0.067 & 0.115 & 0.134 \\
\hline
DiffSR (ours) & 3.81 & \textbf{0.244} & \textbf{0.835} & \textbf{0.479} & \textbf{0.556} & \textbf{0.611} & \textbf{0.221} & \textbf{0.351} & \textbf{0.441} & \textbf{0.091} & \textbf{0.203} & \textbf{0.277} \\
\hline

\end{tabular}
\end{table*}

\textbf{Denoiser.} Given the reliable condition image $y_i'$, we then leverage a conditional diffusion model for our generation module (GM). The primary unit of the diffusion model is the denoiser. At time step $t$, the denoiser receives the output $y_{i,t}$ from the previous denoiser, $y_i'$ from the first stage, and the time embedding $T_{emb}(t)$ as inputs. These inputs are processed to extract relevant features in the image space, producing a prediction $y_{i,t-1}$.
We employ an encoder-decoder network as the architecture of the denoiser~\cite{saharia2022image}.

 \begin{figure}[t]
\centerline{\includegraphics[width=0.51\textwidth]{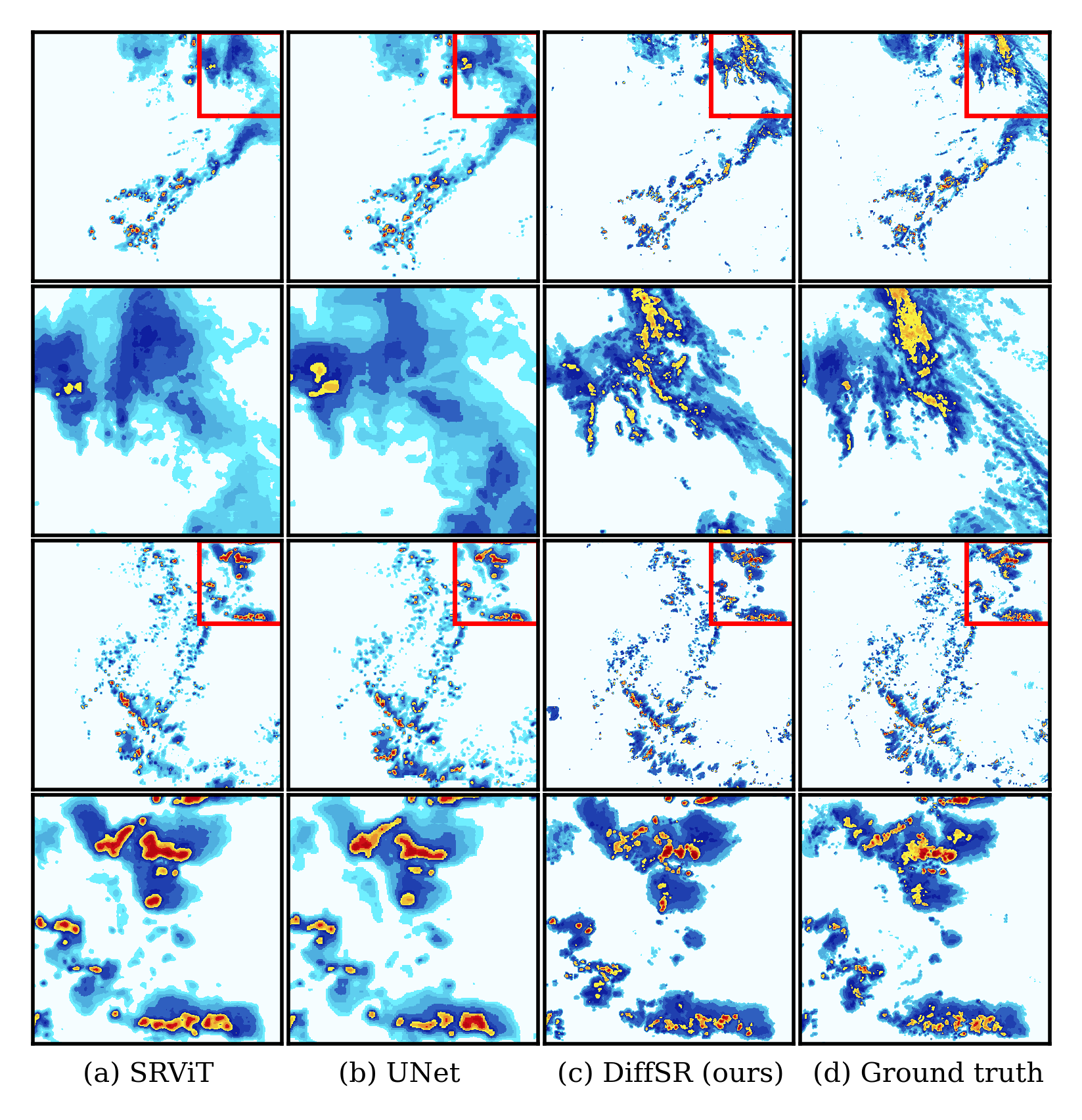}}
\caption{Reconstruction visualization of different models. The results from SRViT and UNet lack details at smaller scales. In contrast, our DiffSR is capable of reconstructing detailed information effectively.}
\label{viz1}
\end{figure}

 \begin{figure}[t]
\centerline{\includegraphics[width=0.49\textwidth]{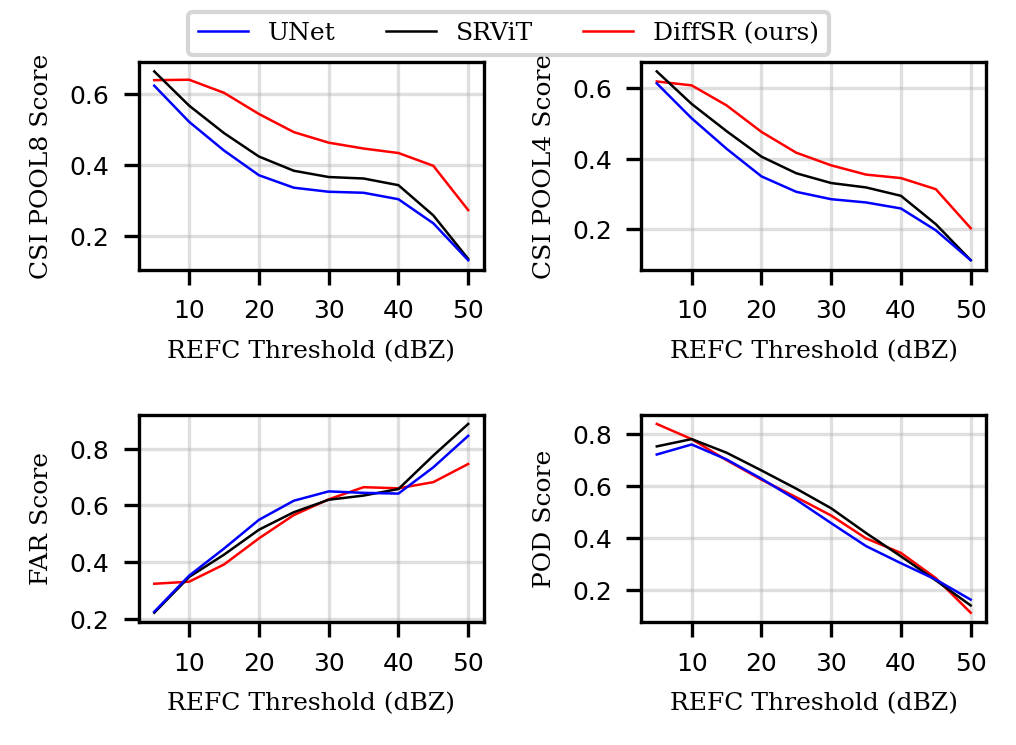}}
\caption{Classification metrics at different composite reflectivity thresholds for DiffSR, SRViT and UNet.}
\label{categorical metrics}
\end{figure}
\section{Experiments}
\subsection{Implementation Details}
\subsubsection{Dataset}We train DiffSR based on the dataset from \cite{hilburn2023gremlin2020, hilburn2023gremlin2021, hilburn2023gremlin2022} over the  contiguous
United States (CONUS). The study uses Advanced Baseline Imager (ABI) infrared Channels 7, 9, 13 and a Geostationary Lightning Mapper (GLM) from the  Geostationary Operational Environmental Satellite (GOES) as input, with composite radar reflectivity from Multi-Radar Multi-Sensor (MRMS) product as target. 

The original image size is 768$\times$1536 pixels, with a 3-kilometer spatial resolution and a 15-minute temporal resolution. However, due to the slow inference speed of diffusion models with such large dimensions, we divide each image into 18 patches of 256$\times$256 using a sliding window approach. We set a threshold $\gamma$ and discard patches that have fewer than $\gamma$ pixels exceeding a value of 6 to avoid data skew. 
We use data from April to September 2020-2021 and April to July 2022 to construct the training dataset. For testing, we use data from August to September 2022. 
\subsubsection{Implementation}
We train DiffSR for 180 epochs with a total batch size of 256 and a threshold of $\gamma=1000$ on 8$\times$80GB NVIDIA A100 GPUs. The model was optimized using the Adam optimizer, with mean squared error (MSE) as the loss function. The learning rate is set to 1e-4. we employ SRViT \cite{stock2024srvit} as our transformation module, which contains a patches-based embedding and positional embedding to convert the input to a feature map with positional information. Then, the feature vector is processed through transformer operations to generate the results for the first stage.

\subsubsection{Metrics} For evaluation, we use traditional classification metrics, including Probability of Detection (POD), False Alarm Ratio (FAR), and Critical Success Index (CSI). Root-Mean-Squared Error (RMSE) is used as a regression metric to measure the difference between ground truth and prediction. To assess perceptual similarity, we also employ Learned Perceptual Image Patch Similarity (LPIPS), and Structural Similarity Index Measure (SSIM).

\subsubsection{Baselines} In our study, we compare the performance of our proposed model with established baseline architectures.
\cite{hilburn2020development} employs an encoder-decoder architecture, which we refer to as ``UNet'' in our study. \cite{stock2024srvit} utilizes a patch-based embedding followed by a transformer architecture, denoted as ``SRViT'' in our work. UNet serves as our baseline for performance comparison, enabling us to evaluate the effectiveness of our proposed approach relative to existing methods.

\subsection{Experimental Results}
% DiffSR exhibits superior overall performance across multiple evaluation metrics.
DiffSR demonstrates superior performance in both qualitative and quantitative results.

\textbf{DiffSR achieves state-of-the-art results on multiple quantitative metrics, particularly in high-value areas.} We employ multiple thresholds (i.e., 15, 35, 50) across different pools (i.e., pool1, pool4, pool8) of CSI to provide a more multifaceted quantitative evaluation. As shown in Table~\ref{comparison}, DiffSR significantly improves quantitative performance on perceptual metrics such as CSI, SSIM, and LPIPS, even with a slight trade-off in RMSE. It is particularly noteworthy that the largest improvements are seen in CSI-50, suggesting that DiffSR outperforms previous methods in effectiveness for high-value data associated with convective weather events.

\textbf{DiffSR demonstrates the power ability to generate results with both high-value and fine-grained characteristics.} In the first two rows of Fig.~\ref{viz1}, DiffSR generates high-value areas that UNet and SRViT fail to produce. In the last two rows, it is observed that DiffSR generates more detailed features in high-value regions and more accurate contours in low-value areas. 
These observations are aligned with the perpetual quantitative results in Table~\ref{comparison}, which shows our DiffSR achieves the highest performance on LPIPS, SSIM, and CSI with POOL8. 

\begin{figure}[t]
\centerline{\includegraphics{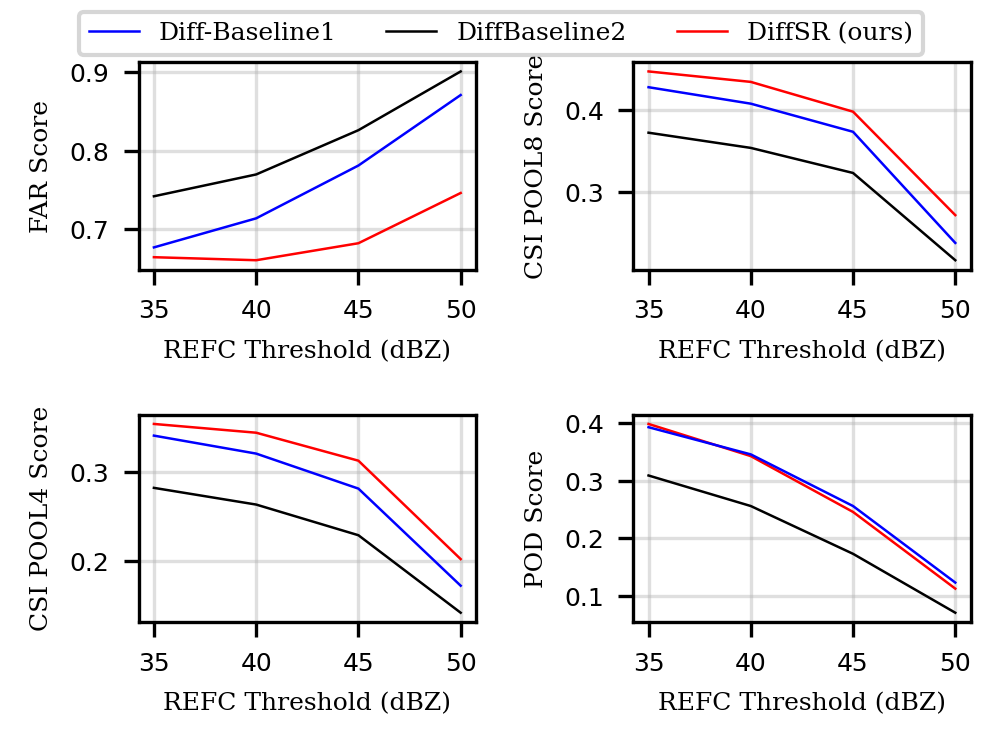}}
\caption{Classification metrics at thresholds 35, 40, 45, 50 for DiffSR, Diff-Baseline1, Diff-Baseline2 model at 500,000th iteration. DiffSR performs better in FAR, CSI POOL8, CSI POOL4.}
\label{compare2}
\end{figure}

\begin{table}[t]
\centering
\caption{Ablation study on condition strategy}
\label{compare_table2}
\begin{tabular}{c|c|c|c|cc}
\hline
\multicolumn{2}{c|}{\textbf{Condition}} &
\multirow{2}{*}{\textbf{$\uparrow$ SSIM}} &
\multirow{2}{*}{\textbf{$\downarrow$ RMSE}} & 
\multicolumn{2}{c}{\textbf{POOL8}}  \\
\cline{1-2}
\cline{5-6}
Satellite & Radar Est. & & & CSI-35 & CSI-50 \\
\hline
$\sqrt{}$& & 0.832 & \textbf{3.77} & 0.427 & 0.238 \\
& $\sqrt{}$ & 0.814 & 4.32 & 0.372 & 0.218 \\
\hline
$\sqrt{}$ & $\sqrt{}$ & \textbf{0.835} & 3.81 & \textbf{0.441} & \textbf{0.277}    \\
\hline
\end{tabular}

\end{table}
\begin{figure}[t]
\centerline{\includegraphics{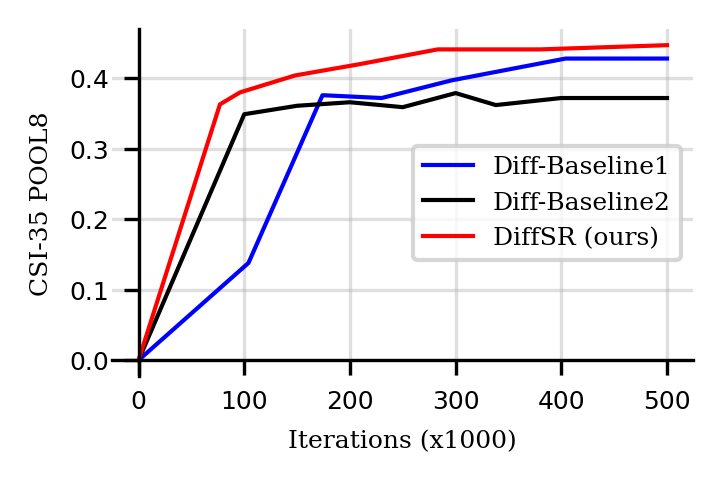}}
\caption{CSI-35 POOL8 for DiffSR, diff-baseline1, and diff-baseline2 model.}
\label{learn}
\end{figure}

\subsection{Ablation Studies and Analysis}

We conduct ablation studies on the conditional inputs of diffusion models, analyzing the convergence of the model under different conditions. In Table~\ref{compare_table2}, Diff-baseline1 utilizes satellite data as the condition of the diffusion model, while Diff-baseline2 relies solely on radar estimation.

\textbf{Effectiveness of satellite data condition}. Using satellite data solely as a condition for the diffusion model results in insufficient accuracy despite yielding stronger perceptual generation outcomes. Table~\ref{compare_table2} reveals that while Diff-baseline1 achieves higher LPIPS, its performance on CSI-50 is significantly inferior to our method, which is highly unfavorable to the observation of convective weather.

\textbf{Effectiveness of radar estimation condition.} As shown in Table~\ref{compare_table2}, the overall performance of Diff-baseline2 is limited when using radar estimation data alone as a condition for the diffusion model due to cumulative errors.
Although radar estimation already contains the contours and high-value areas of the observational data, it also brings certain estimation errors. 
This situation leads to error accumulation and limited performance improvement when enhancing details using only radar estimation data as conditions.

\textbf{Convergence of diffusion models.} In Fig.~\ref{learn}, the training process of diff-baseline1, which uses only satellite data conditions, is the least stable and has the slowest convergence rate. Diff-baseline2, which uses radar estimation conditions, converges more rapidly but performs worse. 
By leveraging a two-stage framework, we first use ViT to generate the radar estimation on a image-level. Then, we combine the satellite data and radar estimation as conditions for the diffusion model. In Figure~\ref{learn}, DiffSR achieves superior results in both convergence speed and performance.

\section{Conclusion}
We propose a universal diffusion-based framework named DiffSR, which leverages satellite data to generate radar composite reflectivity with finer details. The framework consists of two stages: modal transformation and detail generation. Extensive experiments demonstrate that DiffSR achieves SOTA performance on multiple
quantitative metrics. Furthermore, our DiffSR exhibits the power ability to generate results with both high-value and fine-grained characteristics. 
Even though DiffSR incorporates multiple conditions to speed up convergence in the training stage, it still demands a considerable number of steps to yield results during the inference stage, which is computationally expensive.
% However, despite its superior performance, the slower inference speed of the diffusion model leads to significantly higher consumption of computational resources compared to other models. 
% Additionally, the optimization of conditioning mechanisms remains an area for improvement. 
% Future work could address these gaps by incorporating cross-attention mechanisms to better manage different conditions, mapping the image space to the latent space via an encoder, and employing inference techniques similar to \cite{song2020denoising} to accelerate inference speed.

\bibliographystyle{ieeetr}
\bibliography{references}

\end{document}